\documentclass[pra,aps,eqsecnum,amsmath,amssymb,showpacs,showkeys,nofootinbib]{revtex4}
\usepackage[mathscr]{eucal}
\usepackage{xtheorem}

\newcommand{\pbi}{\pmb{\mathbb{I}}}
\newcommand{\wrho}{\widetilde{\rho}}
\newcommand{\wta}{\widetilde{\mathsf{A}}}
\newcommand{\wtb}{\widetilde{\mathsf{B}}}
\newcommand{\wtc}{\widetilde{\mathsf{C}}}
\newcommand{\wtd}{\widetilde{\mathsf{D}}}
\newcommand{\wtu}{\widetilde{\mathsf{U}}}
\newcommand{\wtv}{\widetilde{\mathsf{V}}}
\newcommand{\wpo}{\widetilde{p}}

\newtheorem{plain}{Thm}{Theorem}[section]
{Lemma}
{Definition}

\begin{document}
\clearpage
\preprint{}

\title{Bounds on Shannon distinguishability in terms of partitioned measures}
\author{Alexey E. Rastegin}
\affiliation{A. E. Rastegin, Department of Theoretical Physics, Irkutsk State University,
Gagarin Bv. 20, Irkutsk 664003, Russia}

\begin{abstract}
A family of quantum measures like the Shannon distinguishability
is presented. These measures are defined over the two classes of
POVM measurements and related to separate parts in the expression
for mutual information. Changes of Ky Fan's norms and the
partitioned trace distances under the operation of partial trace
are discussed. Upper and lower bounds on the introduced quantities
are obtained in terms of partitioned trace distances and
Uhlmann's partial fidelities. These inequalities provide a kind of
generalization of the well-known bounds on the Shannon
distinguishability. The notion of cryptographic exponential
indistinguishability for quantum states is revisited. When
exponentially fast convergence is required, all the metrics
induced by unitarily invariant norms are shown to be equivalent.
\end{abstract}

\maketitle

\pagenumbering{arabic}
\setcounter{page}{1}

\section{Introduction}

The advances of last decades have already shown a potential power of
quantum systems as tools for information processing. In all the
topics we deal with information by encoding symbols into quantum
states. Any decoding procedure is done by quantum measurement at
some stage. Because the outcomes of quantum measurement are not
deterministic inevitably, convenient criteria of
distinguishability for measurement statistics are of great
importance. Classical probability distributions as well as mixed
quantum states can be compared in many different ways
\cite{graaf,nielsen05,hayashi}. In the cryptographic context, the
Shannon distinguishability has been found to be very valuable
\cite{graaf}. For example, security of quantum key distribution
against wide classes of attacks has been stated with use of just
this measure \cite{biham,boykin}.

In general, numerous approaches to measuring informational content
of quantum states and their closeness have been developed
\cite{hayashi}. Some of these measures are related to frequently
used norms on the state space \cite{watrous1}. For instance, the
trace norm distance is basic in many issues of quantum
information. However, this measure is not monotone under taking
tensor powers of density operators. Such a monotonicity is
provided by the quantum fidelity elaborated by Uhlmann
\cite{uhlmann76}. In Refs. \cite{jozsa94,caves} a physical meaning
of the fidelity was developed. Though both the trace distance and
the fidelity are most important, more specialized measures can be
needed with respect to the subject. For certain applications, some
of them are more appropriate than others. So, the sub-fidelity
\cite{uhlmann09} and the super-fidelity \cite{uhlmann09,foster}
have been proposed as those measures that are easier to compute.
In effect, many useful relations between various distances are
known \cite{graaf,hayashi}. Further, the Shannon
distinguishability, the trace distance and the fidelity can be
found to be equivalent in posing the exponential
indistinguishability of protocols families \cite{graaf}. Thus, studies of
distinguishability measures and relations between them are still
an actual issue of quantum information theory. The aim of the present
work is to obtain more detailed characterization for Shannon
distinguishability in a refined scale.

\section{Notation and background}

In this section, we recall the notion of Shannon
distinguishability as well as needed facts from matrix analysis.
Let $B$ and $X$ be two random variables assigned to the input and
output of a communication channel. Their probability distributions
$p(b)$ and $p(x)$ are marginal with respect to the joint
probability distribution $p(b,x)$, i.e.
\begin{equation}
p(b)=\sum\nolimits_{x\in{X}}p(b,x) \ ,
\qquad p(x)=\sum\nolimits_{b\in{B}}p(b,x)
\ . \label{bxmarg}
\end{equation}
The relation $p_{b}(x){\,}p(b)=p(b,x)=p_{x}(b){\,}p(x)$ gives the
conditional probabilities $p_{b}(x)$ and $p_{x}(b)$. In terms of
the Shannon entropy, the {\it mutual information} is defined as
\begin{equation}
I(B;X)\triangleq H(B)+H(X)-H(B,X)
\ , \label{nutinf}
\end{equation}
where the joint entropy $H(B,X)=-\sum{\,}{p(b,x)}{\,}\log{p(b,x)}$ and logarithms
are taken to base two. The measure (\ref{nutinf}) quantifies how the
joint distribution $p(b,x)$ differs from the product of marginal
distributions \cite{hayashi}. If we define the entropy of $B$ conditional
on knowing $X$,
\begin{equation}
H(B|X)= - \sum\nolimits_{b\in{B},x\in{X}}{p(b,x)}{\,}\log{p_x(b)}
\ , \label{conbx}
\end{equation}
and also the conditional entropy of $X$ similarly, then
$I(B;X)=H(B)-H(B|X)=H(X)-H(X|B)$ \cite{hayashi}. So, the mutual
information expresses the decrease of uncertainty through the
detection, when uncertainty is quantified by the Shannon entropy.
It is handy to use the binary entropy function
$h(p)\equiv-p{\,}\log{p}-(1-p){\,}\log(1-p)$ and the function
$J(p)\equiv{1-h(p)}$. For a binary input $B$ with equal prior
probabilities \cite{graaf,biham},
\begin{equation}
I(B;X)=\sum\nolimits_{x\in{X}}p(x){\,}J\bigl(p_{x}(0)\bigr)=\sum\nolimits_{x\in{X}}p(x){\,}J\bigl(p_{x}(1)\bigr)
\ , \label{ijbx}
\end{equation}
where $2p(x)=p_{0}(x)+p_{1}(x)$,
$p_{x}(0)=p_{0}(x)/\bigl(2p(x)\bigr)$ and
$p_{x}(1)=p_{1}(x)/\bigl(2p(x)\bigr)$. So, distinguishing the
input is reduced to distinguishing $p_{0}(x)$ and $p_{1}(x)$
\cite{graaf,biham}. Thinking of the expression (\ref{ijbx}) as a
function of the two probability distributions $p_{0}(x)$ and
$p_{1}(x)$, we define the {\it Shannon distinguishability} between
them as
\begin{equation}
SD\bigl(p_{0}(x),p_{1}(x)\bigr)\triangleq
I(B;X)=\sum\nolimits_{x\in{X}}p(x){\,}J\bigl(p_{x}(0)\bigr)
\ . \label{sdclas}
\end{equation}
It is symmetric in the arguments, nonnegative and bounded from
above by the inequality
$SD\bigl(p_{0}(x),p_{1}(x)\bigr)\leq{D}\bigl(p_{0}(x),p_{1}(x)\bigr)\equiv(1/2)\sum_{x\in{X}}|p_{0}(x)-p_{1}(x)|$.

A general quantum measurement is described by {\it ''positive
operator-valued measure''}. The POVM
${\mathcal{A}}=\{{\mathsf{A}}_x\}$ ($x\in{X}$) is a set of
positive matrices obeying $\sum_{x\in{X}}{\mathsf{A}}_x={\pbi}$,
where ${\pbi}$ is the identity in $d$-dimensional Hilbert space
${\mathcal{H}}$ \cite{hayashi}. When the property
${\mathsf{A}}_x{\,}{\mathsf{A}}_y=\delta_{xy}{\,}{\mathsf{A}}_x$
additionally holds, we have a standard measurement described by
{\it ''projector-valued measure''} (PVM). Applying the POVM
${\mathcal{A}}=\{{\mathsf{A}}_x\}$ to a system in the state
$\rho_i$ results in the probability distribution
$p_i^{\cal{A}}(x)={\rm{Tr}}(\rho_i{\mathsf{A}}_x)$. The quantity
\begin{equation}
{\rm{SD}}^{\cal{A}}(\rho_0,\rho_1)\triangleq{SD}\bigl(p_0^{\cal{A}}(x),p_1^{\cal{A}}(x)\bigr)
\label{sdabq}
\end{equation}
shows a distinguishability of the equiprobable states ${\rho}_0$
and ${\rho}_1$ once a particular POVM is used. The {\it Shannon
distinguishability} between the two density matrices is then
defined by \cite{graaf,biham}
\begin{equation}
{\rm{SD}}(\rho_0,\rho_1)\triangleq{\sup}\bigl\{{\rm{SD}}^{\cal{A}}(\rho_0,\rho_1):{\>}{\cal{A}}\in{\rm{POVMs}}\bigr\}
\ , \label{sdabqq}
\end{equation}
where the supremum is taken over all POVMs. This quantity
expresses the amount of information gained in performing a
measurement. No analytic formula for ${\rm{SD}}(\rho_0,\rho_1)$
solely in terms of ${\rho}_0$ and ${\rho}_1$ is known
\cite{graaf}. It is for this reason that easily computable bounds
are desired, particularly in cryptographic applications
\cite{biham,boykin}. Let ${\wrho}_0$ and ${\wrho}_1$ be two
density operators defined on the tensor product
${\mathcal{G}}\otimes{\mathcal{H}}$. The Shannon
distinguishability cannot increase under operation of partial
trace, that is \cite{biham}
\begin{equation}
{\rm{SD}}(\rho_0,\rho_1)\leq{\rm{SD}}({\wrho}_0,{\wrho}_1)
\ , \label{fbouns}
\end{equation}
where the reduced operators
$\rho_i={\rm{Tr}}_{\mathcal{G}}({\wrho}_i)$ are obtained by
tracing-out $N$-dimensional space ${\mathcal{G}}$. The second
upper bound is very important. Let $|{\mathsf{A}}|$ denote a
unique positive square root of
${\mathsf{A}}^{\dagger}{\mathsf{A}}$. For any two density
operators $\rho_0$ and $\rho_1$, there holds \cite{graaf}
\begin{equation}
{\rm{SD}}(\rho_0,\rho_1)\leq \frac{1}{2}{\>}{\rm{Tr}}|\rho_0-\rho_1|\equiv{\rm{D}}_{\rm{tr}}(\rho_0,\rho_1)
\ . \label{sbouns}
\end{equation}
The  upper bounds (\ref{fbouns}) and (\ref{sbouns}) are regularly
used in analysis of vulnerability of quantum key distribution
\cite{biham,boykin}. We will also use both the lower bounds in
terms of the quantum fidelity and the probability of error. The
fidelity between density matrices $\rho_0$ and $\rho_1$ is defined
as ${\rm{F}}(\rho_0,\rho_1)={\rm{Tr}}|\sqrt{\rho_0}\sqrt{\rho_1}|$
\cite{hayashi,watrous1}. The {\it probability of error} between
two probability distributions is given by
$PE\bigl(p_0(x),p_1(x)\bigr)\equiv(1/2)\sum\nolimits_{x\in{X}}\min\{p_0(x),p_1(x)\}$
\cite{graaf}. Minimizing
$PE\bigl(p_0^{\cal{A}}(x),p_1^{\cal{A}}(x)\bigr)$ over all
measurements, the probability of error between $\rho_0$ and
$\rho_1$ is obtained. This task occurs in the problem of state
discrimination \cite{helstrom}. For two equiprobable states, we
have \cite{graaf}
\begin{equation}
{\rm{PE}}(\rho_0,\rho_1)=\frac{1}{2}{\>}\bigl(1-{\rm{D}}_{\rm{tr}}(\rho_0,\rho_1)\bigr)
\ . \label{pepval}
\end{equation}
This value is actually reached by a PVM. The lower bounds on the
Shannon distinguishability are then expressed as \cite{graaf}
\begin{align}
1-{\rm{F}}_0(\rho_0,\rho_1)&\leq{\rm{SD}}(\rho_0,\rho_1)
\ , \label{lowff}\\
J\bigl({\rm{PE}}(\rho_0,\rho_1)\bigr)&\leq{\rm{SD}}(\rho_0,\rho_1)
\ . \label{lowpp}
\end{align}

Below some results of linear algebra will be needed. A unitarily
invariant norm, in signs $|||\centerdot|||$, is a norm on square
matrices that enjoys
$|||{\mathsf{A}}|||=|||{\mathsf{U}}{\mathsf{A}}{\mathsf{V}}|||$
for any ${\mathsf{A}}$ and all unitary ${\mathsf{U}}$,
${\mathsf{V}}$ \cite{bhatia}. Two classes of such norms are
specially important. For real $q\geq1$, the {\it Schatten
$q$-norm} of operator ${\mathsf{A}}$ on ${\mathcal{H}}$ is defined
by
$\|{\mathsf{A}}\|_q=\bigl(\sum\nolimits_{x=1}^{d}s_x({\mathsf{A}})^q
\bigr)^{1/q}$ \cite{watrous1,bhatia}, where the {\it singular
values} $s_x({\mathsf{A}})$ are eigenvalues of $|{\mathsf{A}}|$.
This class includes the trace norm $\|{\mathsf{A}}\|_{\rm{tr}}$
for $q=1$, the Frobenius norm $\|{\mathsf{A}}\|_F$ for $q=2$, and
the spectral norm $\|{\mathsf{A}}\|_{\infty}$ for $q\to\infty$
\cite{watrous1}. The Schatten norms have found use in various
questions of quantum information theory \cite{watrous2}. For
$k=1,\ldots,d$, the {\it Ky Fan $k$-norm} $\|\mathsf{A}\|_{(k)}$
is defined as the sum of $k$ largest singular values
\cite{bhatia}. We obtain the spectral norm for $k=1$ and the trace
norm for $k=d$. We will also use Ky Fan's maximum principle
\cite{kyfan} which can be expressed as follows. If the eigenvalues $\lambda_x$ of Hermitian
operator ${\mathsf{A}}$ are so arranged that
$\lambda_1\geq\lambda_2\geq\cdots\geq\lambda_d{\,}$, then
\begin{equation}
\sum\nolimits_{x=1}^{k} \lambda_x=
\max\bigl\{{\,}{\rm{Tr}}({\mathsf{\Pi}}{\mathsf{A}}):
{\>}{\mathbf{0}}\leq{\mathsf{\Pi}}\leq{\pbi},{\>}{\rm{Tr}}({\mathsf{\Pi}})={k}\bigr\}
\ , \label{kfmaxpr02}
\end{equation}
where the maximization is over positive matrices ${\mathsf{\Pi}}$ with trace $k$ that satisfy
${\mathsf{\Pi}}\leq{\pbi}$.

\section{Definitions of partitioned measures}

In this section, the definitions of new distinguishability
measures are given. For obtaining a more thorough description,
separate terms in the entry for mutual information should be
estimated. That is, we are interested in weight of separate
components in the right-hand side of (\ref{sdclas}). This can be
attained by consideration of partial sums under the decreasing
order of summands. Let ${\#}(X)$ denote the cardinality of the set
$X$.

\begin{Def}
The $k$-th partial Shannon
distinguishability between two the probability distributions
$p_{0}(x)$ and $p_{1}(x)$ is defined by
\begin{equation}
SD_{k}\bigl(p_{0}(x),p_{1}(x)\bigr)\triangleq \max\left\{
\sum\nolimits_{x\in{Y}}p(x){\,}J\bigl(p_{x}(0)\bigr):{\>}Y\subset{X},\ {\#}(Y)=k\right\}
\ . \label{classdk}
\end{equation}
\end{Def}

We have $SD_{l}\bigl(p_{0}(x),p_{1}(x)\bigr)\leq
{SD}_{k}\bigl(p_{0}(x),p_{1}(x)\bigr)$ whenever $l\leq{k}$, and
the entry ${\#}(Y)=k$ can be replaced by ${\#}(Y)\leq{k}$. For the
two probability distributions, we obtain a family of ${\#}(X)$
nonnegative symmetric measures which are all bounded. Let us
proceed to the case of quantum system with the state space
${\mathcal{H}}$. For given POVM ${\cal{A}}$, $k$-th partial
Shannon distinguishability between $\rho_0$ and $\rho_1$ is
naturally put as
\begin{equation}
{\rm{SD}}_k^{\cal{A}}(\rho_0,\rho_1)={SD}_k\bigl(p_0^{\cal{A}}(x),p_1^{\cal{A}}(x)\bigr)
\ . \label{qusdk}
\end{equation}
Further, it is not insignificant that a family of utilized
measurements may be constrained in some ways. Restrictions can be
due to used apparatus, applied protocol or strategy, and perhaps
{\it a priori} information on the signal quantum states. So, it is
of some interest to consider specialized classes of POVM
measurements.

\begin{Def}
Let ${\mathfrak{S}}$ be a family of
POVMs. Then the $k$-th partial Shannon distinguishability with
respect to ${\mathfrak{S}}$ is defined by
\begin{equation}
{\rm{SD}}_k^{\mathfrak{S}}(\rho_0,\rho_1)\triangleq
\sup\left\{{\rm{SD}}_k^{\cal{A}}(\rho_0,\rho_1):{\>}{\cal{A}}\in{\mathfrak{S}}\right\}
\ . \label{defsdfr}
\end{equation}
\end{Def}

In the following, we will consider the two important families of
POVMs. Putting $d={\rm{dim}}({\mathcal{H}})$, the first family
${\mathfrak{A}}$ is defined as
\begin{equation}
{\mathfrak{A}}\triangleq\left\{{\cal{A}}:{\>}{\rm{Tr}}({\mathsf{A}}_x)\leq{1} {\>\>}\forall{\>\>} x\in{X},
\ {\#}(X)\leq d^2\right\}
\ . \label{deffpa}
\end{equation}
Indispensable one-rank POVMs are all contained in this family. As
a rule, quantum information tasks lead to hard problems of
nonlinear optimization. Due to famous Davies' results
\cite{davies}, an analysis can often be simplified to a POVM with
one-rank elements whose number is limited by
$d\leq{\#}(X)\leq{d}^2$. Using this fact, Fuchs and Peres have
shown that the optimal detection for a two-state system is reached
with a two-dimensional eavesdropper's probe \cite{peresf}. Such
POVMs are sufficient for optimal unambiguous discrimination
\cite{peres2} which is widely adopted in quantum key distribution
\cite{brandt05}. We also know that POVMs with elements of higher
rank can never give more mutual information than maximizing
one-rank POVM. So, the family ${\mathfrak{A}}$ of measurements is
of importance. The second family ${\mathfrak{B}}$ is defined as
\begin{equation}
{\mathfrak{B}}\triangleq\bigl\{{\cal{A}}:{\>}{\rm{Tr}}({\mathsf{A}}_x)\geq{1}{\>\>}\forall{\>\>}
x\in{X}\bigr\} \ . \label{deffpb}
\end{equation}
In this definition, we have ${\#}(X)\leq{d}$ with necessity. The
family ${\mathfrak{B}}$ contains all the projective measurements
which are easier to realize experimentally. Moreover, in
discrimination between two quantum states the average probability
of error is minimized by POVM that is actually a PVM
\cite{helstrom}.

In addition, reasons for using the families ${\mathfrak{A}}$ and
${\mathfrak{B}}$ are connected with interpretations of the
partitioned trace distances and the partial fidelities in terms of
measurement statistics. Such relations between classical
distinguishability measures and their quantum versions are used in
various contexts \cite{hayashi,caves}. The {\it $k$-th partitioned
trace distance} between  ${\rho}_0$ and ${\rho}_1$ is expressed by
\cite{rast091}
\begin{equation}
{\rm{D}}_k(\rho_0,\rho_1)=\frac{1}{2}{\>}\|\rho_0-\rho_1\|_{(k)}
\ . \label{defpard2}
\end{equation}
For $k=d$, this definition leads to the trace norm distance
${\rm{D}}_{\rm{tr}}(\rho_0,\rho_1)$ which can also be put via
extremal properties of quantum operations \cite{rast07}. The
partitioned distances enjoy many properties of the trace norm
distance, including the unitary invariance and the strong
convexity \cite{rast091}. The derivation of these results is
essentially based on the Ky Fan maximum principle
(\ref{kfmaxpr02}). Let us put the {\it $k$-th classical trace
distance} between probability distributions $p_{0}(x)$ and
$p_{1}(x)$ as
\begin{equation}
D_{k}\bigl(p_{0}(x),p_{1}(x)\bigr)\triangleq\max\left\{
\frac{1}{2}\sum\nolimits_{x\in{Y}}|p_{0}(x)-p_{1}(x)|:{\>}Y\subset{X},\ {\#}(Y)=k\right\}
\ . \label{cladk}
\end{equation}
A kind of statistical interpretation is then expressed by
\begin{equation}
{\rm{D}}_k(\rho_0,\rho_1)=\max\left\{D_k\bigl(p_0^{\cal{A}}(x),p_1^{\cal{A}}(x)\bigr):
{\>}{\cal{A}}\in{\mathfrak{A}}\right\}
\ , \label{qcreldis}
\end{equation}
where the maximum is actually reached by the one-rank PVM \cite{rast091}.

In Ref. \cite{uhlmann00}, Uhlmann introduced {\it $k$-th partial
fidelity} as
\begin{equation}
{\rm{F}}_k(\rho_0,\rho_1)=\min\left\{\sum\nolimits_{x\in{Y}}
s_x(\sqrt{\rho_0}\sqrt{\rho_1}):\ {\#}(Y)=d-k\right\}
\ . \label{depfid1}
\end{equation}
These quantities allow to resolve the equivalence of pairs of
mixed states under invertible transformations \cite{uhlmann00}.
The partial fidelities enjoy the joint concavity \cite{uhlmann00}
and other useful properties \cite{rast092}. For $k=0$, we have the
regular quantum fidelity
${\rm{F}}_0(\rho_0,\rho_1)={\rm{Tr}}|\sqrt{\rho_0}\sqrt{\rho_1}|$.
Note that
${\rm{F}}_k(\rho_0,\rho_1)\equiv{\rm{F}}_0(\rho_0,\rho_1)-\|\sqrt{\rho_0}\sqrt{\rho_1}\|_{(k)}$
in terms of the Ky Fan $k$-norm. We also define the $k$-th
fidelity between probability distributions as \cite{rast092}
\begin{equation}
F_k\bigl(p_{0}(x),p_{1}(x)\bigr)\triangleq\min\biggl\{\sum_{{\,}x\in{Y}}
\sqrt{p_{0}(x)p_{1}(x)}:{\>}Y\subset{X},\ {\#}(Y)={\#}(X)-k\biggr\}
\ . \label{depfid2}
\end{equation}
A kind of statistical interpretation of the partial fidelities is
provided by \cite{rast092}
\begin{equation}
{\rm{F}}_k(\rho_0,\rho_1)\leq\inf\left\{F_k\bigl(p_0^{\cal{A}}(x),p_1^{\cal{A}}(x)\bigr):
{\>}{\cal{A}}\in{\mathfrak{B}}\right\}
\ , \label{qcrelfid}
\end{equation}
where the equality cannot always be reached in general.

\section{Changes under the operation of partial trace}

In quantum theory, the state of a subsystem of a composite quantum
system is described by a reduced density operator. Except for the
opaque method, for attack on a quantum cryptosystem the intruder
entangles his probes with transmitted carriers somehow
\cite{brandt03}. In either case, the intruder finally deals with
those density matrices that are results of the operation of
partial trace. Hence we are interested in how used quantitative
measures may be changed by this operation. Let ${\wrho}_0$ and
${\wrho}_1$ be density matrices on
${\mathcal{G}}\otimes{\mathcal{H}}$ and
$\rho_i={\rm{Tr}}_{{\mathcal{G}}}({\wrho}_i)$. For the partial
fidelities there holds \cite{rast092}
\begin{equation}
{\rm{F}}_k(\rho_0,\rho_1)
\geq {\rm{F}}_{(Nk)}({\wrho}_0,{\wrho}_1)
\ . \label{geqpart}
\end{equation}
We shall now give a similar relation for partitioned trace
distances. For distances between the marginal probability
distributions $p_i(x)=\sum_{1\leq\xi\leq{N}}\wpo_i(\xi,x)$, we
have
\begin{equation}
D_k\bigl(p_{0}(x),p_{1}(x)\bigr)=\sum_{x\in{Y'}}
{\,}\biggl|{\,}\sum_{\xi=1}^{N}\wpo_0(\xi,x)-\sum_{\xi=1}^{N}\wpo_1(\xi,x){\,}\biggr|
\leq D_{(kN)}\bigl(\wpo_0(\xi,x),\wpo_1(\xi,x)\bigr)
{\>} \label{theo33p1}
\end{equation}
due to the triangle inequality. Here $Y'$ denotes a $k$-subset of
$X$ such that the maximum in (\ref{cladk}) is reached. Using the
statistical interpretation (\ref{qcreldis}), we could obtain a
quantum version of (\ref{theo33p1}). However, it is of some
interest to consider more general question about the Ky Fan norms.
In Ref. \cite{lidar1} the problem is resolved for those unitarily
invariant norms that are multiplicative over tensor products. In
particular, there hold
\begin{equation}
\|{\mathsf{A}}\|_F\leq\sqrt{N}{\,}\|{\wta}\|_F \ ,
\qquad \|{\mathsf{A}}\|_{\infty}\leq{N}{\,}\|{\wta}\|_{\infty}
\ , \label{fropnorm}
\end{equation}
where ${\mathsf{A}}$ is taken from $\wta$ by the operation of
partial trace over $N$-dimensional space. Except for $k=1$, however, Ky Fan's norms
are not multiplicative in this way \cite{lidar1}. So the following
result is valuable.

\begin{Lem}\label{kfchan}
For each operator $\wta$ on the
tensor product ${\mathcal{G}}\otimes{\mathcal{H}}$,
${\rm{dim}}({\mathcal{G}})=N$, and its partial trace
${\mathsf{A}}={\rm{Tr}}_{\mathcal{G}}({\wta})$, there holds
\begin{equation}
\|{\mathsf{A}}\|_{(k)}\leq\|{\wta}\|_{(Nk)}
\ . \label{akakn}
\end{equation}
\end{Lem}

{\bf Proof.} (a) Let ${\wta}$ be Hermitian. If
${\wta}={\wtb}-{\wtc}$ is the Jordan decomposition, then both
${\mathsf{B}}\equiv{\rm{Tr}}_{\mathcal{G}}({\wtb})$,
${\mathsf{C}}\equiv{\rm{Tr}}_{\mathcal{G}}({\wtc})$ are positive
(but not mutually orthogonal in general) and
${\mathsf{A}}={\mathsf{B}}-{\mathsf{C}}$. Since ${\mathsf{A}}$ is also Hermitian,
there holds
$\|{\mathsf{A}}\|_{(k)}={\rm{Tr}}\bigl[({\mathsf{P}}-{\mathsf{Q}}){\,}{\mathsf{A}}\bigr]$
for some mutually orthogonal projectors ${\mathsf{P}}$ and
${\mathsf{Q}}$ that satisfy
${\rm{rank}}({\mathsf{P}}+{\mathsf{Q}})\leq{k}$ \cite{rast091}. Due to positivity
of ${\mathsf{P}}$, ${\mathsf{Q}}$, ${\mathsf{B}}$, and
${\mathsf{C}}$,
\begin{align}
\|{\mathsf{A}}\|_{(k)}&={\rm{Tr}}_{\mathcal{H}}\bigl[({\mathsf{P}}-{\mathsf{Q}}){\,}{\mathsf{A}}\bigr]
={\rm{Tr}}_{\mathcal{H}}\bigl[({\mathsf{P}}-{\mathsf{Q}})({\mathsf{B}}-{\mathsf{C}})\bigr]
\nonumber\\
&\leq{\rm{Tr}}_{\mathcal{H}}\bigl[({\mathsf{P}}+{\mathsf{Q}})({\mathsf{B}}+{\mathsf{C}})\bigr]=
{\rm{Tr}}_{{\mathcal{G}}\otimes{\mathcal{H}}}\bigl[({\pbi}_{\mathcal{G}}\otimes{\mathsf{\Pi}})
({\wtb}+{\wtc})\bigr]
\ , \label{ppabc2}
\end{align}
where we used ${\mathsf{\Pi}}={\mathsf{P}}+{\mathsf{Q}}$ and the
properties of the trace including \cite{watrous1}
\begin{equation}
{\rm{Tr}}_{\mathcal{H}}({\mathsf{\Pi}}{\,}{\mathsf{B}})=
{\rm{Tr}}_{{\mathcal{G}}\otimes{\mathcal{H}}}\bigl[({\pbi}_{\mathcal{G}}\otimes{\mathsf{\Pi}})
{\,}{\wtb}{\,}\bigr]
\ . \label{ppabc3}
\end{equation}
Since ${\wtb}+{\wtc}=|{\wta}|$ by definition, its eigenvalues are
positive and
${\rm{rank}}({\pbi}_{\mathcal{G}}\otimes{\mathsf{\Pi}})\leq{Nk}$,
Ky Fan's maximum principle (\ref{kfmaxpr02}) provides the relation
(\ref{akakn}) for the Hermitian case.

(b)\footnote {An extension to the non-Hermitian case was noted by
one of the referees.} For arbitrary ${\wta}$, we define its left
absolute value
$|{\wta}|_L=\bigl({\wta}{\,}{\wta}^{\dagger}\bigr)^{1/2}$ and
right absolute value
$|{\wta}|_R=\bigl({\wta}^{\dagger}{\wta}\bigr)^{1/2}$. It follows
from Hermiticity of these two operators and part (a) that the
Ky Fan $(Nk)$-norm satisfies
\begin{equation}
\|{\wta}\|_{(Nk)}=\||{\wta}|_L\|_{(Nk)}=\||{\wta}|_R\|_{(Nk)}
\geq\max\left\{\|{\mathsf{A}}_L\|_{(k)},\|{\mathsf{A}}_R\|_{(k)}\right\}
\ , \label{lrwta11}
\end{equation}
where
${\mathsf{A}}_L\equiv{\rm{Tr}}_{\mathcal{G}}\bigl(|{\wta}|_L\bigr)$
and
${\mathsf{A}}_R\equiv{\rm{Tr}}_{\mathcal{G}}\bigl(|{\wta}|_R\bigr)$.
We claim that the right-hand side of (\ref{lrwta11}) is not less
than $\|{\mathsf{A}}\|_{(k)}$. Using the singular value
decomposition ${\wta}={\wtu}{\,}{\wtd}{\,}{\wtv}$, we obtain
$|{\wta}|_L={\wtu}{\,}{\wtd}{\,}{\wtu}^{\dagger}$ and
$|{\wta}|_R={\wtv}^{\dagger}{\wtd}{\,}{\wtv}$. To prove the claim,
we write
\begin{equation}
{\wtu}=\sum_{i,j=1}^N |i\rangle\langle j|\otimes{\mathsf{U}}_{ij}
\>, \quad {\wtd}=\sum_{j=1}^N |j\rangle\langle j|\otimes{\mathsf{D}}_{jj}
\>, \quad {\wtv}=\sum_{i,j=1}^N |i\rangle\langle j|\otimes{\mathsf{V}}_{ij}
\>, \label{wudv}
\end{equation}
with respect to an orthonormal basis $\{|i\rangle\}$ in the space
${\mathcal{G}}$. That is, ${\wtu}$ may be viewed as a $N$-by-$N$
block matrix with blocks ${\mathsf{U}}_{ij}$, and so on. By
calculations, we get
\begin{equation}
{\mathsf{A}}=\sum_{i,j=1}^N {\mathsf{U}}_{ij}{\mathsf{D}}_{jj}{\mathsf{V}}_{ji}
\>, \quad {\mathsf{A}}_L=\sum_{i,j=1}^N {\mathsf{U}}_{ij}{\mathsf{D}}_{jj}{\mathsf{U}}_{ji}^{{\,}\dagger}
\>, \quad {\mathsf{A}}_R=\sum_{i,j=1}^N {\mathsf{V}}_{ij}^{{\,}\dagger}{\mathsf{D}}_{jj}{\mathsf{V}}_{ji}
\>, \label{aalaar}
\end{equation}
because
${\rm{Tr}}_{\mathcal{G}}\bigl(|i\rangle\langle{j}|\otimes{\mathsf{\Theta}}\bigr)=\langle{j}|i\rangle{\,}{\mathsf{\Theta}}$
for any operator ${\mathsf{\Theta}}$ on ${\mathcal{H}}$. Let us
use the two 1-by-$N^2$ block matrices
\begin{align}
{\mathsf{L}}&=
\left[{\>}{\mathsf{U}}_{11}\sqrt{\mathsf{D}}_{11} {\quad} {\mathsf{U}}_{12}\sqrt{\mathsf{D}}_{22}
{\quad} \cdots {\quad} {\mathsf{U}}_{NN}\sqrt{\mathsf{D}}_{NN}
{\>}\right] \ , \label{lldef}\\
{\mathsf{R}}&=
\left[{\>}{\mathsf{V}}_{11}^{{\,}\dagger}\sqrt{\mathsf{D}}_{11}
{\quad} {\mathsf{V}}_{12}^{{\,}\dagger}\sqrt{\mathsf{D}}_{22}
{\quad} \cdots {\quad}
{\mathsf{V}}_{NN}^{{\,}\dagger}\sqrt{\mathsf{D}}_{NN} {\>}\right]
\ . \label{rrdef}
\end{align}
It is easy to check that
${\mathsf{A}}={\mathsf{L}}{\mathsf{R}}^{\dagger}$,
${\mathsf{A}}_L={\mathsf{L}}{\mathsf{L}}^{\dagger}$ and
${\mathsf{A}}_R={\mathsf{R}}{\mathsf{R}}^{\dagger}$. We finally
have
\begin{equation}
\|{\mathsf{L}}{\mathsf{R}}^{\dagger}\|_{(k)}\leq \left(\|{\mathsf{L}}{\mathsf{L}}^{\dagger}\|_{(k)}\|{\mathsf{R}}{\mathsf{R}}^{\dagger}\|_{(k)}\right)^{1/2}
\leq\max\left\{\|{\mathsf{L}}{\mathsf{L}}^{\dagger}\|_{(k)},\|{\mathsf{R}}{\mathsf{R}}^{\dagger}\|_{(k)}\right\}
\ . \label{prfin}
\end{equation}
The inequality on the left is a Cauchy-Schwarz inequality for
ordinary products of rectangular matrices (of the same size) and
given unitarily invariant norms (see, e.g., the inequality
(3.5.22) in \cite{hjhon91}). $\blacksquare$

In particular, we have
$\|{\mathsf{A}}\|_{\rm{tr}}\leq\|{\wta}\|_{\rm{tr}}$ for $k=d$,
$\|{\mathsf{A}}\|_{\infty}\leq\|{\wta}\|_{(N)}$ for $k=1$. The
latter relation is stronger than the second inequality of
(\ref{fropnorm}) (except when the largest singular value of
${\wta}$ has multiplicity $\geq{N}$). The partitioned trace
distances satisfy
\begin{equation}
{\rm{D}}_k(\rho_0,\rho_1)\leq {\rm{D}}_{(Nk)}({\wrho}_0,{\wrho}_1)
\ . \label{corl33}
\end{equation}
As it is well known, the trace norm distance cannot increase and
the fidelity cannot decrease under the partial trace. This
endorses the mind reason that objects become less distinguishable
when only partial information is available. All the partitioned
distances enjoy the mentioned property in the sense of relations
(\ref{theo33p1}) and (\ref{corl33}). For Uhlmann's partial
fidelities, the relation (\ref{geqpart}) is useful in this regard.

\section{Basic inequalities}

Because the Shannon distinguishability measures
${\rm{SD}}_k^{\mathfrak{S}}(\rho_0,\rho_1)$ are positive-valued,
we are foremost interested in upper bounds similar to
(\ref{fbouns}) and (\ref{sbouns}). First, we present the
inequalities with the partitioned trace distances. Corresponding
bounds for density operators are essentially based on the
relations for probability distributions.

\begin{Thm}\label{sddf}
Let the measures $SD_{k}$, $D_{k}$, and
$F_{k}$ be defined by formulas (\ref{classdk}), (\ref{cladk}), and
(\ref{depfid2}) respectively. For any two probability
distributions and $k=0,1,\ldots,{\#}(X)$,
\begin{equation}
SD_{k}\bigl(p_{0}(x),p_{1}(x)\bigr)\leq
D_{k}\bigl(p_{0}(x),p_{1}(x)\bigr)\leq1-F_{k}\bigl(p_{0}(x),p_{1}(x)\bigr)
\ . \label{lem31}
\end{equation}
\end{Thm}

{\bf Proof.} Firstly, we denote by $Y'$ a $k$-subset of $X$ such
that
$$
SD_{k}\bigl(p_{0}(x),p_{1}(x)\bigr)=\sum\nolimits_{x\in{Y'}}p(x){\,}J\bigl(p_{x}(0)\bigr) \ .
$$
As it is shown in \cite{graaf,biham}, for $0\leq{r}\leq{1}$ there
holds $J(r)\leq|2r-1|$, whence
\begin{equation}
SD_{k}\bigl(p_{0}(x),p_{1}(x)\bigr)\leq\sum_{x\in{Y'}}p(x)
\left|\bigl(p_{0}(x)/p(x)\bigr)-1\right|=\frac{1}{2}\sum_{x\in{Y'}}
\left|p_{0}(x)-p_{1}(x)\right|
\label{lem311}
\end{equation}
due to $p(x)=\bigl(p_{0}(x)+p_{1}(x)\bigr)/2$. By (\ref{cladk}),
the right-hand side of (\ref{lem311}) does not exceed $k$-th
partitioned distance $D_{k}\bigl(p_{0}(x),p_{1}(x)\bigr)$.
Secondly, let $Y\subset{X}$ be a $k$-subset such that the maximum
in (\ref{cladk}) is reached. Because $\sum_{x\in{X}}p_i(x)=1$, we
write
\begin{align}
2{\,}D_{k}\bigl(p_{0}(x),p_{1}(x)\bigr)&\leq\sum_{x\in{Y}} \bigl(p_{0}(x)+p_{1}(x)\bigr)
= 2{\,}- \sum_{x\in{Y_c}}\bigl(p_{0}(x)+p_{1}(x)\bigr) \nonumber\\
& \leq {\,}2{\,}- \sum_{x\in{Y_c}}2{\,}\sqrt{p_{0}(x)p_{1}(x)}\leq{\,} 2 - 2{\,}F_{k}\bigl(p_{0}(x),p_{1}(x)\bigr)
\ , \label{lem312}
\end{align}
where $Y_c$ is the complement of $Y$ and, therefore, $\#(Y_c)=\#(X)-k$. $\blacksquare$

Due to (\ref{qcreldis}) and (\ref{lem31}), for all ${\cal{A}}\in{\mathfrak{A}}$
we have ${\rm{SD}}_k^{\cal{A}}(\rho_0,\rho_1)\leq{\rm{D}}_k(\rho_0,\rho_1)$.
If each number of the set is not greater than ${\rm{D}}_k(\rho_0,\rho_1)$ then
the supremum of the set does also obey this. Combining the claim
with (\ref{corl33}), we obtain an extension of the upper bounds
(\ref{fbouns}) and (\ref{sbouns}) in terms of partitioned measures.

\begin{Thm}\label{sdkaa}
Let the measures ${\rm{SD}}_k^{\mathfrak{A}}$ and ${\rm{D}}_k$ be defined by
formulas (\ref{defsdfr}), for the family (\ref{deffpa}), and
(\ref{defpard2}) respectively. For any two density matrices and
$k=0,1,\ldots,d$,
\begin{equation}
{\rm{SD}}_k^{\mathfrak{A}}(\rho_0,\rho_1)\leq{\rm{D}}_k(\rho_0,\rho_1)
\ . \label{theo32}
\end{equation}
{\it If operators $\rho_0$ and $\rho_1$ are taken as
$\rho_i={\rm{Tr}}_{{\mathcal{G}}}({\wrho}_i)$ over $N$-dimensional
space ${\mathcal{G}}$ then}
\begin{equation}
{\rm{SD}}_k^{\mathfrak{A}}(\rho_0,\rho_1)\leq{\rm{D}}_{(kN)}({\wrho}_0,{\wrho}_1)
\ . \label{theo32w}
\end{equation}
\end{Thm}

The inequality (\ref{theo32}) generalizes the well-known bound
(\ref{sbouns}) to the case considered. In analysis of quantum
information protocols, the operation of partial trace is
inevitable. Apparently, no simple version of (\ref{fbouns}) exists
for partial measures ${\rm{SD}}_k^{\mathfrak{A}}(\rho_0,\rho_1)$.
But the bound (\ref{fbouns}) is rather useful in a ready
combination with (\ref{sbouns}), namely
\begin{equation}
{\rm{SD}}(\rho_0,\rho_1)\leq {\rm{D}}_{\rm{tr}}({\wrho}_0,{\wrho}_1)
\ . \label{sbounc}
\end{equation}
Indeed, the Shannon distinguishability itself is typically
unknown. For instance, in a study of security problem the result
(\ref{sbounc}) is actually used \cite{biham,boykin}. So, a useful
analog of (\ref{sbounc}) is provided by (\ref{theo32w}). Let us
proceed to the relations with the partial fidelities.

\begin{Thm}\label{sdkbb}
Let the measures
${\rm{SD}}_k^{\mathfrak{B}}$ and ${\rm{F}}_k$ be defined by
formulas (\ref{defsdfr}), for the family (\ref{deffpb}), and
(\ref{depfid1}) respectively. For any two density matrices and
$k=0,1,\ldots,d$,
\begin{equation}
{\rm{SD}}_k^{\mathfrak{B}}(\rho_0,\rho_1)\leq1-{\rm{F}}_k(\rho_0,\rho_1)
\ . \label{theo33}
\end{equation}
{\it If operators $\rho_0$ and $\rho_1$ are taken as
$\rho_i={\rm{Tr}}_{{\mathcal{G}}}({\wrho}_i)$ over $N$-dimensional
space ${\mathcal{G}}$ then}
\begin{equation}
{\rm{SD}}_k^{\mathfrak{B}}(\rho_0,\rho_1)\leq{\rm{SD}}_{k}^{\widetilde{\mathfrak{B}}}({\wrho}_0,{\wrho}_1)
\ . \label{theo33w}
\end{equation}
\end{Thm}

{\bf Proof.} By (\ref{lem31}) and (\ref{qcrelfid}), we get
${\rm{SD}}_k^{\cal{A}}(\rho_0,\rho_1)\leq1-{\rm{F}}_k(\rho_0,\rho_1)$
for any POVM ${\cal{A}}\in{\mathfrak{B}}$. Combining this with the
definition (\ref{defsdfr}) at once gives (\ref{theo33}). Further,
the set
$\widetilde{\cal{A}}=\{{\pbi}_{\mathcal{G}}\otimes{\mathsf{A}}_x\}$
is a POVM on the total space ${\mathcal{G}}\otimes{\mathcal{H}}$
and generates the probabilities
\begin{equation}
{\rm{Tr}}_{\mathcal{H}}(\rho_i{\,}{\mathsf{A}}_x)={\rm{Tr}}_{{\mathcal{G}}\otimes{\mathcal{H}}}
\bigl[{\wrho}_i({\pbi}_{\mathcal{G}}\otimes{\mathsf{A}}_x)\bigr] \ ,
\end{equation}
whence
${\rm{SD}}_k^{\cal{A}}(\rho_0,\rho_1)={\rm{SD}}_{k}^{\widetilde{\cal{A}}}({\wrho}_0,{\wrho}_1)$.
The fact
${\rm{Tr}}_{{\mathcal{G}}\otimes{\mathcal{H}}}({\pbi}_{\mathcal{G}}\otimes{\mathsf{A}}_x)\geq{N}$
implies $\widetilde{\cal{A}}\in\widetilde{\mathfrak{B}}$. So the
left-hand side of (\ref{theo33w}) is the supremum over a certain
subfamily of $\widetilde{\mathfrak{B}}$. $\blacksquare$

Note that the combined relation
${\rm{SD}}_k^{\mathfrak{B}}(\rho_0,\rho_1)\leq1-{\rm{F}}_{k}({\wrho}_0,{\wrho}_1)$
may rather be suitable in calculations. Both the basic
inequalities (\ref{theo32}) and (\ref{theo33}) can be posed as
majorization relations. Notions of majorization theory are very
useful, for instance, in matrix analysis \cite{bhatia} and studies
of quantum systems \cite{vidal}. Let $q=(q_1,\ldots,q_m)$ and
$r=(r_1,\ldots,r_m)$ be elements of real space ${\mathbb{R}}^m$.
We say that $q$ is {\it weakly submajorized} by $r$, in symbols
$q\prec_w r$, when \cite{bhatia}
\begin{equation}
\sum\nolimits_{x=1}^{k} q_x^{\downarrow} \leq
\sum\nolimits_{x=1}^{k} r_x^{\downarrow} \ ,\quad 1\leq{k}\leq{m}
\ , \label{major}
\end{equation}
where the arrows down indicate that vector coordinates are put in
decreasing order. Denoting
$p_i^{\cal{A}}(x)={\rm{Tr}}\bigl(\rho_i{\,}{\mathsf{A}}_x\bigr)$
and
$J_x^{\cal{A}}={J}\bigl(p_x^{\cal{A}}(0)\bigr)\equiv{J}\bigl(p_x^{\cal{A}}(1)\bigr)$,
the inequalities (\ref{theo32}) and (\ref{theo33}) are merely
reformulated as
\begin{align}
& \bigl(p_0^{\cal{A}}+p_1^{\cal{A}}\bigr){\,}J^{\cal{A}} \pmb{\prec_w} s(\rho_0-\rho_1)
\quad\forall{\>\>}{\cal{A}}\in{\mathfrak{A}}
\ , \label{maj1}\\
& \bigl(p_0^{\cal{A}}+p_1^{\cal{A}}\bigr){\,}J^{\cal{A}} \pmb{\prec_w} 2s\bigl(\sqrt{\rho_0}\sqrt{\rho_1}\bigr)
\quad\forall{\>\>}{\cal{A}}\in{\mathfrak{B}}
\ , \label{maj2}
\end{align}
where
$p_x^{\cal{A}}(i)=p_i^{\cal{A}}(x)/\bigl(p_0^{\cal{A}}(x)+p_1^{\cal{A}}(x)\bigr)$
and the definitions (\ref{defpard2}) and (\ref{depfid1}) are
expanded. The majorization relations (\ref{maj1}) and (\ref{maj2})
give another description for components of Shannon
distinguishability measures in more detailed terms. In a certain
sense, these statements are complementary to each other, since
they are related to the two different families of practically
important POVM measurements. The following bounds are analogs of
(\ref{lowff}) and (\ref{lowpp}) for the partial Shannon
distinguishability measures.

\begin{Thm}\label{sdkcc} 
Let the measures ${\rm{SD}}_k^{\mathfrak{A}}$ and ${\rm{SD}}_k^{\mathfrak{B}}$ be
defined by (\ref{defsdfr}), for the classes (\ref{deffpa}) and
(\ref{deffpb}), ${\rm{F}}_0$ by (\ref{depfid1}) and ${\rm{PE}}$ by
(\ref{pepval}). For $k=0,1,\ldots,d$, there hold
\begin{align}
\frac{k}{d^2}{\,}\bigl(1-{\rm{F}}_0(\rho_0,\rho_1)\bigr)&\leq{\rm{SD}}_k^{\mathfrak{A}}(\rho_0,\rho_1)
\ , \label{lowbaa}\\
\frac{k}{d}{\>}J\bigl({\rm{PE}}(\rho_0,\rho_1)\bigr)&\leq{\rm{SD}}_k^{\mathfrak{B}}(\rho_0,\rho_1)
\ . \label{lowbbb}
\end{align}
\end{Thm}

{\bf Proof.} If we put the partial sums $Q_k=\sum_{x=1}^k
q_x^{\downarrow}$, then (see lemma 3 in \cite{rast092})
\begin{equation}
mQ_k\geq{k}Q_m \ , \quad k=0,1,\ldots,m
\ . \label{ll44}
\end{equation}
Due to this relation and the condition $\#(X)\leq{d}^2$ in (\ref{deffpa}), for each
$\cal{A}\in{\mathfrak{A}}$ we have
\begin{equation}
{\rm{SD}}_k^{\cal{A}}(\rho_0,\rho_1)\geq\frac{k}{d^2}{\>}{\rm{SD}}^{\cal{A}}(\rho_0,\rho_1)
\ . \label{sdksdd}
\end{equation}
Hence the suprema of the two sides of (\ref{sdksdd}) satisfy
${\rm{SD}}_k^{\mathfrak{A}}(\rho_0,\rho_1)\geq(k/d^2){\,}{\rm{SD}}^{\mathfrak{A}}(\rho_0,\rho_1)$.
The measure ${\rm{SD}}^{\mathfrak{A}}(\rho_0,\rho_1)$ is the
Shannon distinguishability (\ref{sdabqq}) itself, as the family
${\mathfrak{A}}$ certainly contains one-rank POVM that optimizes
the mutual information. So the bound (\ref{lowbaa}) follows from
(\ref{lowff}). Second, let $\varPi\in{\mathfrak{B}}$ be
PVM such that ${\rm{PE}}(\rho_0,\rho_1)=PE\bigl(p_0^{\varPi}(x),p_1^{\varPi}(x)\bigr)$.
Using (\ref{ll44}) and the definitions (\ref{defsdfr}) and
(\ref{deffpb}), we have
\begin{equation}
{\rm{SD}}_k^{\mathfrak{B}}(\rho_0,\rho_1)\geq
{\rm{SD}}_k^{\varPi}(\rho_0,\rho_1)\geq\frac{k}{d}{\>\,}{\rm{SD}}^{\varPi}(\rho_0,\rho_1)
\ . \label{sdkspp}
\end{equation}
From the relation
$SD\bigl(p_0^{\varPi}(x),p_1^{\varPi}(x)\bigr)\geq{\,}J\bigl(PE\bigl(p_0^{\varPi}(x),p_1^{\varPi}(x)\bigr)\bigr)$,
which is known for probability distributions \cite{graaf}, we
obtain (\ref{lowbbb}). $\blacksquare$

The significance of Theorems \ref{sdkaa}, \ref{sdkbb} and \ref{sdkcc} is that, while the
quantum Shannon distinguishability measures are unknown in a
closed form, the inequalities provide a useful way to estimate
them. A more detailed characterization is given with respect to
those POVMs that are important from the practical viewpoint. Both
the partitioned trace distances and partial fidelities enjoy a
kind of statistical interpretation. On the other hand, they do not
have a direct information-theoretic meaning. Such a treatment may
be expressed via the partial varieties of Shannon
distinguishability. Due to the lower bounds (\ref{lowbaa}) and
(\ref{lowbbb}), partitioned measures can also be applied in the
context of exponential indistinguishability.

\section{Notes on exponential indistinguishability}

Comparing the protocol implementation (i.e. the family of
protocols) with the ideal protocol specification, we would like
that the probability of cheating for each participant vanishes
exponentially, as taken {\it security parameter} $n$ increases
\cite{graaf}. This label may sign the length of a string, the
number of rounds, or the number of carriers transmitted. Let
$\bigl\{X_0\bigr\}=\bigl\{{X}_0^{(1)},{X}_0^{(2)},{X}_0^{(3)},\ldots\bigr\}$
and
$\bigl\{X_1\bigr\}=\bigl\{{X}_1^{(1)},{X}_1^{(2)},{X}_1^{(3)},\ldots\bigr\}$
be families of random variables with the probability distributions
$\bigl\{{p}_0^{(1)},{p}_0^{(2)},{p}_0^{(3)},\ldots\bigr\}$ and
$\bigl\{{p}_1^{(1)},{p}_1^{(2)},{p}_1^{(3)},\ldots\bigr\}$. These
families are {\it exponentially indistinguishable} if there exist
some $n_0$ and $\varepsilon\in(0;1)$ such that \cite{graaf}
\begin{equation}
D\bigl(p_0^{(n)},p_1^{(n)}\bigr)\leq \varepsilon^n \quad\forall{\>\>}n\geq n_0
\ . \label{exindef}
\end{equation}
The motivation and examples are presented in \cite{graaf}. The
measures $D(p_0,p_1)$, $PE(p_0,p_1)$, $F_0(p_0,p_1)$ and
$SD(p_0,p_1)$ are found to be equivalent when we require
exponentially fast convergence to the values that are obtained for
two identical distributions (i.e., $D=0$, $PE=1/2$, $F_0=1$, and
$SD=0$). It is natural to take two families
$\bigl\{\rho_0^{(n)}\bigr\}=\bigl\{\rho_0^{(1)},\rho_0^{(2)},\rho_0^{(3)},\ldots\bigr\}$
and
$\bigl\{\rho_1^{(n)}\bigr\}=\bigl\{\rho_1^{(1)},\rho_1^{(2)},\rho_1^{(3)},\ldots\bigr\}$
of density operators on $d$-dimensional space $\mathcal{H}$. The
two families are exponentially indistinguishable if there exist
some $n_0$ and $\varepsilon\in(0;1)$ such that \cite{graaf}
\begin{equation}
{\rm{D}}_{\rm{tr}}\bigl(\rho_0^{(n)},\rho_1^{(n)}\bigr)\leq \varepsilon^n \quad\forall{\>\>}n\geq{n_0}
\ . \label{exindefqq}
\end{equation}
It is valuable that an equivalence of similar kind takes place in
the quantum case. Namely, an exponentially fast convergence with
respect to one of the measures ${\rm{D}}$, ${\rm{PE}}$,
${\rm{F}}_0$ and ${\rm{SD}}$ implies the same with respect to all these measures \cite{graaf}.
Below, we will analyze a convergence with respect to both the partitioned
trace distances and partial varieties of Shannon
distinguishability.

\begin{Thm}\label{exin1} Let $\{{\rm{M}}_k\}$ be one of three
measure series $\bigl\{{\rm{D}}_k\bigr\}_{k=1}^{d}$,
$\bigl\{{\rm{SD}}_k^{\mathfrak{A}}\bigr\}_{k=1}^{d^2}$, and
$\bigl\{{\rm{SD}}_k^{\mathfrak{B}}\bigr\}_{k=1}^{d}$ defined by
formulas (\ref{defpard2}) and (\ref{defsdfr}), for the classes
(\ref{deffpa}) and (\ref{deffpb}), respectively. If families
$\bigl\{\rho_0^{(n)}\bigr\}$ and $\bigl\{\rho_1^{(n)}\bigr\}$ are
exponentially indistinguishable with respect to measure
${\rm{M}}_{k_0}$ of series $\{{\rm{M}}_k\}$ then they are
exponentially indistinguishable with respect to all measures of
the series.
\end{Thm}

{\bf Proof.} (a) Suppose the families $\bigl\{\rho_0^{(n)}\bigr\}$
and $\bigl\{\rho_1^{(n)}\bigr\}$ are exponentially
indistinguishable with respect to ${\rm{D}}_{k_0}$
($1\leq{k_0}\leq{d}$). So there exist integer $n_0$ and real
$\varepsilon\in(0;1)$ such that
\begin{equation}
{\rm{D}}_{k_0}\bigl(\rho_0^{(n)},\rho_1^{(n)}\bigr)\leq \varepsilon^n \quad\forall{\>\>}n\geq{n_0}
\ . \label{exindk}
\end{equation}
Due to (\ref{ll44}), the trace norm distance obeys
${\rm{D}}_{\rm{tr}}\bigl(\rho_0^{(n)},\rho_1^{(n)}\bigr)\leq(d/k_0){\,}\varepsilon^n$,
whenever $n\geq{n_0}$. Let $n_{\varepsilon}$ denote the smallest
integer such that
\begin{equation}
\left(\frac{d}{k_0}\right)^{1/n_{\varepsilon}}\cdot\varepsilon=\epsilon<1
\ . \label{nepsd}
\end{equation}
This value clearly exists because $\varepsilon<1$ and
$(d/k_0)^{1/n}\to1$ in the limit $n\to\infty$. By calculations, we
get
\begin{equation}
n_{\varepsilon}=\left\lfloor\frac{\ln(d/k_0)}{-\ln\varepsilon}\right\rfloor +1
\ . \label{nepsv}
\end{equation}
For all $n\geq\max\{n_0,n_{\varepsilon}\}$, we then obtain
${\rm{D}}_{\rm{tr}}\bigl(\rho_0^{(n)},\rho_1^{(n)}\bigr)\leq\epsilon^n$.
By definition, each partitioned trace distance is not larger than
the trace norm distance. This completes the proof for the series
$\bigl\{{\rm{D}}_k\bigr\}_{k=1}^{d}$. (b) It follows from
(\ref{sdksdd}) and related reasons that
\begin{equation}
{\rm{SD}}\bigl(\rho_0^{(n)},\rho_1^{(n)}\bigr)\leq\frac{d^2}{k_0}{\>\,}{\rm{SD}}_{k_0}^{\mathfrak{A}}\bigl(\rho_0^{(n)},\rho_1^{(n)}\bigr)
\leq\frac{d^2}{k_0}{\>\,}\varepsilon^n
\label{exinsdk}
\end{equation}
for given $k_0$ and all $n\geq{n_0}$. By the above
arguments, for all $n\geq\max\{n_0,n_{*}\}$ we have
\begin{equation}
{\rm{SD}}\bigl(\rho_0^{(n)},\rho_1^{(n)}\bigr)\leq{\epsilon_{*}}^n
\ , \label{sdepsas}
\end{equation}
where $\epsilon_{*}<1$ and $n_{*}$ are defined by replacing $d$
with $d^2$ in the formulas (\ref{nepsd}) and (\ref{nepsv}). By
definition, each measure ${\rm{SD}}_k^{\mathfrak{A}}$ does not
exceed the total sum ${\rm{SD}}^{\mathfrak{A}}$ and, therefore,
the left-hand side of (\ref{sdepsas}). (c) Suppose that for given
$k_0$ and all $n\geq{n_0}$
\begin{equation}
{\rm{SD}}_{k_0}^{\mathfrak{B}}\bigl(\rho_0^{(n)},\rho_1^{(n)}\bigr)\leq \varepsilon^n
\ . \label{sdbeb}
\end{equation}
Using (\ref{lowbbb}) and the above reasons, there holds
$J{\!}\left({\rm{PE}}\bigl(\rho_0^{(n)},\rho_1^{(n)}\bigr)\right)\leq\epsilon^n$,
whenever $n\geq\max\{n_0,n_{\varepsilon}\}$. By calculus, for
$r\in[0;1]$ we get $J(r)\geq(2/\ln2)(r-1/2)^2$, whence
\begin{equation}
\frac{1}{2}{\,}-{\,}{\rm{PE}}\bigl(\rho_0^{(n)},\rho_1^{(n)}\bigr)\leq
\sqrt{\frac{\ln2}{2}}{\>}\bigl(\sqrt{\epsilon}\bigr)^n
\ . \label{peexid}
\end{equation}
This implies exponentially fast convergence with respect to the
probability of error and, therefore \cite{graaf}, with respect to
the Shannon distinguishability itself. The latter is not less than
the measure
${\rm{SD}}^{\mathfrak{B}}\geq{\rm{SD}}_k^{\mathfrak{B}}$, where
$k=1,\ldots,d$.  $\blacksquare$

Thus, an equivalence stated in the paper \cite{graaf} really is
much more broad in character. Indeed, exponentially
indistinguishable families of density operators enjoy this
property with respect to all the above partitioned measures
(except for the partial fidelities). With respect to the question
of interest, some measures may be easier to calculate or
experimentally estimate. So, a freedom in formulation of
exponential indistinguishability is useful. It turns out that such
a treatment can be proceeded to each metric induced by a unitarily
invariant norm. Due to the Fan dominance theorem (see, e.g., Corollary (3.5.9) in \cite{hjhon91}),
many relations with Ky Fan's norms can be extended to all
unitarily invariant norms. For any traceless Hermitian operator
${\mathsf{A}}$, there hold \cite{eisert}
\begin{equation}
||{\mathsf{A}}||_{\infty}\leq |||{\mathsf{Z}}|||^{-1}|||{\mathsf{A}}||| \ , \qquad
|||{\mathsf{A}}|||\leq\frac{|||{\mathsf{Z}}|||}{2}{\,}||{\mathsf{A}}||_{\rm{tr}}
\ , \label{audes}
\end{equation}
where ${\mathsf{Z}}={\rm{diag}}(1,1,0,\ldots,0)$. Note that the multiplier of $||{\mathsf{A}}||_{\rm{tr}}$ in
(\ref{audes}) is independent of ${\mathsf{A}}$. We say that
$\bigl\{\rho_0^{(n)}\bigr\}$ and $\bigl\{\rho_1^{(n)}\bigr\}$ are
{\it exponentially indistinguishable with respect to the induced
metric} if there exist some $m_0$ and $\delta\in(0;1)$ such that
\begin{equation}
|||\rho_0^{(n)}-\rho_1^{(n)}|||\leq \delta^n \quad\forall{\>\>}n\geq{m_0}
\ . \label{exdefun}
\end{equation}
Since the difference between two density matrices is traceless, we
can use (\ref{audes}). For given unitarily invariant norm, the
value of $|||{\mathsf{Z}}|||$ is a fixed positive number. For the
Schatten $q$-norm, say, $|||{\mathsf{Z}}|||=2^{1/q}$. The claimed
equivalence can be observed in the same manner, as the statement
of Theorem \ref{exin1} has been proved. Using the first inequality of
(\ref{audes}), the formula (\ref{exdefun}) leads to exponentially
fast convergence with respect to the metric $2{\rm{D}}_1$ induced
by the spectral norm. By Theorem \ref{exin1}, the exponentially fast
convergence takes place with respect to all measures of the series
$\bigl\{{\rm{D}}_k\bigr\}_{k=1}^{d}$ including the trace norm
distance. That is, any convergence of a kind (\ref{exdefun}) implies
the convergence of a kind (\ref{exindefqq}). Conversely, the second
inequality of (\ref{audes}) and the formula (\ref{exindefqq}) lead
to $|||\rho_0^{(n)}-\rho_1^{(n)}|||\leq
|||{\mathsf{Z}}|||{\>}\varepsilon^n$, whenever $n\geq{n_0}$. By
some technical work, this implies that the inequality
(\ref{exdefun}) holds for each unitarily invariant norm.

\begin{Thm}\label{exin2} If two families of density matrices are
exponentially indistinguishable with respect to one metric induced
by a unitarily invariant norm then these families are
exponentially indistinguishable with respect to all the metrics
induced by unitarily invariant norms.
\end{Thm}

Thus, unitarily invariant norms provide flexible tools for
analysis of distinguishability including the cryptographic
context. Some of them are very well studied, for instance, the
spectral norm and the trace norm. So their nice properties are
widely adopted in many respects. However, induced metrics do not
have a direct information-theoretic content. This sense is rather
a feature of the Shannon distinguishability and its partial
varieties because they are defined via the mutual information. But
closed analytical expressions for them are not known. Hence all
the above relations between different measures are important.
Moreover, quantum exponential indistinguishability can be resolved
by means of any metric induced by a unitarily invariant norm.

\section{Conclusion}

The partial Shannon distinguishability measures have been
presented. A more detailed characterization is given with respect
to both the adopted measurements and separate terms in the sum for
mutual information. Since the operation of partial trace is
typical, a special issue of Ky Fan's norms after the partial trace was
firstly resolved by the statement of Lemma \ref{kfchan}. In general, the
optimizing measurement can be unknown or infeasible with an
available equipment. So the studied quantities are relevant when
the optimal POVM is replaced by a POVM from the classes
considered. The upper bounds on the introduced measures are given
in a form of simple inequalities using the partitioned trace
distances (see Theorem \ref{sdkaa}) and Uhlmann's partial fidelities (see
Theorem \ref{sdkbb}). In Theorem \ref{sdkcc}, the relevant lower bounds are also
presented. Theorem \ref{exin1} treats the proposed measures in the context
of exponentially indistinguishable families of quantum states. For
such two families, a distinguisher may be unable to identify the
source of a given sample, even if he is not restricted to
polynomial-time calculations. In the case of exponentially fast
convergence, all the metrics induced by unitarily invariant norms
are shown to be tantamount (see Theorem \ref{exin2}). This equivalence is
expected to be useful in designing indistinguishable families of
density matrices.

\acknowledgments

The comments of anonymous referees were very valuable. I am
particularly grateful for pointing out part (b) of the proof
of Lemma \ref{kfchan}.


\begin{thebibliography}{55}

\bibitem{graaf}
Fuchs, C.A., van de Graaf, J.: {\it Cryptographic
distinguishability measures for quantum mechanical states}. IEEE
Trans. Inf. Theory {\bf 45}, 1216--1227 (1999)

\bibitem{nielsen05}
Gilchrist, A., Langford, N.K., Nielsen, M.A.: {\it
Distance measures to compare real and ideal quantum processes}.
Phys. Rev. A {\bf 71}, 062310 (2005)

\bibitem{hayashi}
Hayashi, M.: {\it Quantum Information: An Introduction}. Springer, Berlin (2006)

\bibitem{biham}
Biham, E., Boyer, M., Brassard, G., van de Graaf, J., Mor, T.: {\it
Security of quantum key distribution against all collective
attacks}. Algorithmica {\bf 34}, 372--388 (2002)

\bibitem{boykin}
Biham, E., Boyer, M., Boykin, P.O., Mor, T., Roychowdhury, V.:
{\it A proof of the security of quantum key distribution}. J.
Cryptology {\bf 19}, 381–-439 (2006)

\bibitem{watrous1}
Watrous, J.: {\it CS 798: Theory of quantum information}. University of Waterloo, \\
http://www.cs.uwaterloo.ca/$\sim$watrous/quant-info/lecture-notes/all-lectures.pdf (2008)

\bibitem{uhlmann76}
Uhlmann, A.: {\it The transition probability in the state space of
a *-algebra}. Rep. Math. Phys. {\bf 9}, 273--279 (1976)

\bibitem{jozsa94}
Jozsa, R.: {\it Fidelity for mixed quantum states}. J. Mod. Opt.
{\bf 41}, 2315--2323 (1994)

\bibitem{caves}
Fuchs, C.A., Caves, C.M.: {\it Mathematical techniques for
quantum communication theory}. Open Syst. Inf.
Dyn. {\bf 3}, 345--356 (1995)

\bibitem{uhlmann09}
Miszczak, J.A., Pucha{\l}a, Z., Horodecki, P., Uhlmann, A.,
\.{Z}yczkowski, K.: {\it Sub–- and super–-fidelity as bounds for
quantum fidelity}. Quantum Inf. Comput. {\bf 9}, 0103--0130 (2009)

\bibitem{foster}
Mendon\c{c}a, P.E.M.F., Napolitano, R.d.J., Marchiolli, M.A.,
Foster, C.J., Liang, Y.-C.: {\it Alternative fidelity measure
between quantum states}. Phys. Rev. A {\bf 78}, 052330 (2008)

\bibitem{bhatia}
Bhatia, R.:{\it Matrix Analysis}. Springer, New York (1997)

\bibitem{watrous2}
Watrous, J.: {\it Notes on super-operator norms induced by Schatten
norms}. Quantum Inf. Comput. {\bf 5}, 58–-68 (2005)

\bibitem{kyfan}
Fan, K.: {\it On a theorem of Weyl concerning eigenvalues of linear
transformations. I}. Proc. Nat. Acad. Sci. USA {\bf 35},
652--655 (1949)

\bibitem{davies}
Davies, E.B.: {\it Information and quantum measurement}. IEEE
Trans. Inf. Theory {\bf 24}, 596--599 (1978)

\bibitem{peresf}
Fuchs, C.A., Peres, A.: {\it Quantum-state disturbance versus
information gain: Uncertainty relations for quantum information}.
Phys. Rev. A {\bf 53}, 2038--2045 (1996)

\bibitem{peres2}
Peres, A., Terno, D.R.: {\it Optimal distinction between
non-orthogonal quantum states}. J. Phys. A: Math. Gen. {\bf 31},
7105--7111 (1998)

\bibitem{brandt05}
Brandt, H.E.: {\it Unambiguous state discrimination in quantum key
distribution}. Quantum Inf. Process. {\bf 4}, 387--398 (2005)

\bibitem{helstrom}
Helstrom, C.W.: {\it Quantum Detection and Estimation Theory}.
Academic Press, New York (1976)

\bibitem{rast091}
Rastegin, A.E.: {\it Partitioned trace distances}. Quantum Inf.
Process. {\bf 9}, 61--73 (2010)

\bibitem{rast07}
Rastegin, A.E.: {\it Trace distance from the viewpoint of quantum
operation techniques}. J. Phys. A: Math. Theor. {\bf 40},
9533--9549 (2007)

\bibitem{uhlmann00}
Uhlmann, A.: {\it On ''partial'' fidelities}. Rep. Math. Phys. {\bf 45}, 407--418 (2000)

\bibitem{rast092}
Rastegin, A.E.: {\it Some properties of partial fidelities}.
Quantum Inf. Comput. {\bf 9}, 1069--1080 (2009)

\bibitem{brandt03}
Brandt, H.E.: {\it Optimum probe parameters for entangling probe
in quantum key distribution}. Quantum Inf. Process. {\bf 2},
37--79 (2003)

\bibitem{lidar1}
Lidar, D.A., Zanardi, P., Khodjasteh, K.: {\it Distance bounds on
quantum dynamics}. Phys. Rev. A {\bf 78}, 012308 (2008)

\bibitem{hjhon91}
Horn, R.A. and Johnson, C.R.: {\it Topics in Matrix Analysis}.
Cambridge University Press, Cambridge (1991)

\bibitem{vidal}
Nielsen, M.A., Vidal, G.: {\it Majorization and the
interconversion of bipartite states}. Quantum Inf. Comput. {\bf
1}, 76--93 (2001)

\bibitem{eisert}
Audenaert, K.M.R., Eisert, J.: {\it Continuity bounds on the
quantum relative entropy}. J. Math. Phys. {\bf 46}, 102104 (2005)



\end{thebibliography}
\end{document}